\documentclass[twocolumn]{article}
\usepackage{latex8}
\usepackage{psfig}
\usepackage{rules}
\usepackage{times}
\usepackage{alltt}

\usepackage{url}

\newcommand{\bsl}{\(\backslash\)}

\newcommand\transp{~~~~}

\newcommand\pl[1]{#1}

\newcommand\red[1]{~\ensuremath{\stackrel{\mbox{\tiny #1}}{\longrightarrow}}~}
\newcommand\redr[1]{~\ensuremath{\stackrel{\mbox{\tiny #1}}{\Longrightarrow}}~}
\newcommand\redt[1]{~\ensuremath{\stackrel{\mbox{\tiny #1}}{\longmapsto}}~}
\newcommand\sred{\red{STOP}}
\newcommand\ured{\red{SUSP}}
\newcommand\ered{\red{END}}
\newcommand\ared{\red{$\alpha$}}

\newcommand\uredr{\redr{SUSP}}
\newcommand\eredr{\redr{END}}
\newcommand\aredr{\redr{$\alpha$}}
\newcommand\eredt{\redt{END}}
\newcommand\aredt{\redt{$\alpha$}}

\newcommand\actstop{\mbox{STOP}}
\newcommand\actsusp{\mbox{SUSP}}
\newcommand\actend{\mbox{END}}

\newcommand\tran[3]{#1 #2 #3}
\newcommand\stran[2]{\tran{#1}{\sred}{#2}}
\newcommand\utran[2]{\tran{#1}{\ured}{#2}}
\newcommand\etran[2]{\tran{#1}{\ered}{#2}}
\newcommand\atran[2]{\tran{#1}{\ared}{#2}}
\newcommand\tranr[3]{\tran{#1}{\redr{#2}}{#3}}

\newcommand\utranr[2]{\tran{#1}{\uredr}{#2}}
\newcommand\etranr[2]{\tran{#1}{\eredr}{#2}}
\newcommand\atranr[2]{\tran{#1}{\aredr}{#2}}
\newcommand\etrant[2]{\tran{#1}{\eredt}{#2}}
\newcommand\atrant[2]{\tran{#1}{\aredt}{#2}}

\newcommand\state[3]{#1,#2~#3}
\newcommand\sstate[2]{#1~#2}
\newcommand\st[6]{\stran{\state{#1}{#2}{#3}}{\state{#4}{#5}{#6}}}
\newcommand\ut[6]{\utran{\state{#1}{#2}{#3}}{\state{#4}{#5}{#6}}}
\newcommand\et[6]{\etran{\state{#1}{#2}{#3}}{\state{#4}{#5}{#6}}}
\newcommand\at[6]{\atran{\state{#1}{#2}{#3}}{\state{#4}{#5}{#6}}}
\newcommand\tr[6]{\tranr{\state{#1}{#2}{#3}}{#4}{\sstate{#5}{#6}}}

\newcommand\utr[5]{\utranr{\state{#1}{#2}{#3}}{\sstate{#4}{#5}}}
\newcommand\etr[5]{\etranr{\state{#1}{#2}{#3}}{\sstate{#4}{#5}}}
\newcommand\atr[5]{\atranr{\state{#1}{#2}{#3}}{\sstate{#4}{#5}}}
\newcommand\ett[5]{\etrant{\state{#1}{#2}{#3}}{\sstate{#4}{#5}}}
\newcommand\att[5]{\atrant{\state{#1}{#2}{#3}}{\sstate{#4}{#5}}}

\newcommand\mtt[1]{\mbox{\texttt{#1}}}
\newcommand\fn[2]{\mtt{(fn $#1$ => $#2$)}}
\newcommand\rec[3]{\mtt{(rec $#1(#2)$ => $#3$)}}
\newcommand\true{\mtt{true}}
\newcommand\false{\mtt{false}}
\newcommand\inner[1]{\ensuremath{\langle #1 \rangle}}
\newcommand\cons[2]{\inner{\mtt{#1},#2}}

\begin{document}

\title{\textbf{Reactive Programming in Standard ML}}
\author{Riccardo R. Pucella\\[.1in]
Bell Laboratories\\
Lucent Technologies\\
600 Mountain Avenue\\
Murray Hill, NJ 07974 USA\\
\texttt{riccardo@research.bell-labs.com}}
\maketitle
\thispagestyle{empty}   %

{
\centerline{\normalsize\bfseries Abstract}%
\vspace*{12pt}%
\it%
Reactive systems are systems that maintain an ongoing interaction with
their  environment, activated by receiving input events from the
environment and producing output events in response. Modern
programming languages designed to program such systems use a paradigm
based on the notions of instants and activations. We describe a
library for Standard ML that provides basic primitives for programming
reactive systems.  The library is a low-level system upon which more
sophisticated reactive behaviors can be built, which provides a
convenient framework for prototyping extensions to existing reactive
languages. 
}

\section{Introduction}

We consider in this paper the problem of programming applications
containing reactive subsystems. A reactive system is defined as a
system that maintains an ongoing interaction with its environment
\cite{Harel85}, activated by receiving input events from the
environment and producing output events in response. Typical examples
of reactive systems are user interfaces, required to coordinate the
various user requests (from the keyboard, the mouse and other devices)
with information coming from the application (enabling or disabling
input components and so on). Such systems generally decompose
into independent  parallel components cooperating to solve a given
task, and exhibit a high degree of concurrency
\cite{Gansner92}. Because of this, programming reactive systems using
traditional sequential languages can be difficult, and one often turns
to concurrent languages to simplify the programming task.    

In the past decade, a class of languages has emerged
specifically for programming reactive systems, including imperative
languages such as \pl{Esterel} \cite{Berry92} and declarative languages
such as \pl{Signal} \cite{LeGuernic86} and \pl{Lustre}
\cite{Caspi87}. Those languages are directly based on
the model of reactive systems as being activated by input events and
producing output events. Their approach to programming reactive
systems, referred to as the \emph{reactive paradigm}, is to divide the
life of a reactive system into \emph{instants}, which are the moments
where the system reacts. They allow the programmer to write
statements that depend on instants. For example, a program may wait
for the third instant where a given event occurs, and so on. Instants
provide a notion of \emph{logical time} to which programs may
refer. This is in contrast to languages providing a notion of absolute
(or real) time, for example \pl{Ada} \cite{Ada95} with its
\texttt{delay} statement. 

This approach to programming reactive systems via instants has an
interesting consequence. Instants act as a \emph{global} logical time
for a program, and thus the end of instants provide a 
consistent configuration of the state of the program, where one can
make decisions before the next activation. This in turns allows for a
clean specification of \emph{preemption}, whereby one reaction can
abort another reaction executing in parallel \cite{Berry93}.

This paper describes a library for the programming language
\pl{Standard ML} (\pl{SML}) \cite{Milner97} that implements the
essence of the reactive paradigm, as described by Boussinot
\cite{Boussinot91}: the notions of instants and  activations. It
permits the definition of \pl{SML} expressions that can be activated
and that specify control points denoting the end of instants. 

The original purpose of the library was to help develop a reactive
interface language for connecting user interface components,
independently of the underlying window management system. 
The library also has features that make it interesting in its own
right. It is built from a very small set of primitives, simplifying
the task of analyzing programs using the library. It provides a
framework for reactivity that fits naturally with the
mostly-applicative programming style of \pl{SML}. It can be
implemented without sizable extensions to the language (none if the
implementation provides first-class continuations or a similar
facility). More importantly, it provides an opportunity to study the
interaction of reactive primitives with features missing from most
existing reactive languages: higher-order functions and recursion.  

The library is intended as a low-level framework implementing basic
reactive functionality upon which one may build more sophisticated
machinery. It can be used to investigate and prototype extensions to
existing higher-level reactive languages, extending for example
\pl{Esterel} with higher-order facilities. Compilation for
higher-level reactive languages is non-trivial, and
using the reactive library as a target language for compilation, one
can rapidly prototype extensions. Once a useful extension has been
identified, effort can be put into finding a compilation process that
generates code as efficient as possible.

The paper is organized as follows: the next section describes the primitives
implemented by the library; Section \ref{s:3} provides an example of
reactive code written using the library; Section \ref{s:A} gives the
operational semantics of the reactive primitives; Section \ref{s:4} compares
the library to existing reactive frameworks, and Section \ref{s:5}
concludes with a  discussion of future work.

\section{The reactive library}
\label{s:2}

In this section, we describe the reactive library and provide simple
examples. Primitives for creating and activating basic reactive
expressions are given, along with combinators to create new reactive
expressions by combining existing ones.

\subsection{Basic reactive expressions}

\begin{figure*}[t]
\hrule
\medskip
\begin{alltt}\scriptsize
val rexp        : (unit -> unit) -> rexp    (* create a basic reactive expression        *)

val stop        : unit -> unit              (* stop execution of current instant         *)

val react       : rexp -> bool              (* activate a reactive expression            *)
val reactT      : rexp -> unit              (* activate until termination                *)

val dup         : rexp -> rexp              (* duplicate reactive expression             *)
val activate    : rexp -> unit              (* relinquish control to reactive expression *)
\end{alltt}
\hrule
\caption{Reactive primitives}
\label{f:1}
\end{figure*}

The central notion defined by the reactive library is that of a
\emph{reactive expression}. A reactive expression is fundamentally an
\pl{SML} expression that defines instants. The basic 
primitives are shown in Figure \ref{f:1}. The function \texttt{rexp}
creates a reactive expression out of its argument (a \texttt{unit ->
unit} function). Activating the
reactive expression will evaluate the argument of \texttt{rexp} until
a call to \texttt{stop}, which marks the end of the current
instant. The next activation of the reactive expression will resume
the evaluation from the last point where a \texttt{stop} was
called, until either another \texttt{stop} is called or the evaluation
terminates. As a simple example, consider the following:
\begin{alltt}\scriptsize
val exp = rexp (fn () => (print "FIRST\bsl{}n";
                          stop();
                          print "SECOND\bsl{}n"))
\end{alltt}
This code defines a reactive expression \texttt{exp} that prints
\texttt{FIRST} the first time it is activated, and \texttt{SECOND} the
second time it is activated. After the second activation, the reactive
expression is \emph{terminated}.

To activate a reactive expression from SML code, one applies
the function \texttt{react}, which returns \texttt{true} if the
expression is   terminated and \texttt{false} otherwise. Activating a
terminated expression has no effect. The
function \texttt{dup} creates a copy of the reactive 
expression, with its current state. Here's a
sample session with the above example: 
\begin{alltt}\scriptsize
- react (exp);
FIRST                      (* first instant     *)
false : bool
- val copy = dup (exp);    (* make a copy       *)
val copy = - : rexp
- react (exp);
SECOND                     (* second instant    *)
true : bool                (* rexp terminates   *)
- react (exp);
true : bool
- react (copy);            (* activate the copy *)
SECOND
true : bool
\end{alltt}

The function \texttt{reactT} repeatedly activates the reactive
expression until it terminates. 

A reactive expression can furthermore relinquish control to another
reactive expression, via the function \texttt{activate}. When a
reactive expression calls \texttt{activate} on a reactive expression
$e$, it effectively behaves as $e$ until $e$ terminates, at which
point the reactive expression continues evaluation. For
example, activating the reactive expression
\begin{alltt}\scriptsize
rexp (fn ()=> (activate (exp);
               print "DONE\bsl{}n"))
\end{alltt}
activates \texttt{exp}, stopping when \texttt{exp} stops. Once
\texttt{exp} terminates, evaluation of the reactive expression
continues and \texttt{DONE} is printed before the reactive expression
itself terminates.

\subsection{Combinators}

Reactive expressions with a more complex behavior are created via
combinators, which take existing reactive expressions and produce new
ones. Figure \ref{f:2} presents the most important combinators
implemented by the library.

The combinator \texttt{merge} takes two reactive expressions $e_1$ and
$e_2$ and returns a new reactive expression $e$ with the following
behavior: when $e$ is activated, it activates $e_1$ and then
$e_2$. The reactive expression $e$ terminates when both $e_1$ 
and $e_2$ terminate. For example, consider the reactive expression: 
\begin{alltt}\scriptsize
merge (rexp (fn ()=> (print "1"; stop (); print "2")),
       rexp (fn ()=> (print "A"; stop (); print "B")))
\end{alltt}
This reactive expression will print \texttt{1A} at the first instant,
and \texttt{2B} at the second, then terminate. Note that the order of
activation of the branches of the \texttt{merge} is determined. This
ensures deterministic evaluation of the reactive
expression. Alternate orders of evaluation can be specified by
defining micro-instants (see Section \ref{s:2.4}).

The combinator \texttt{rif} takes a boolean-valued function and two
reactive expressions $e_1$ and $e_2$, and returns a new reactive
expression $e$ with the following behavior: when $e$ is activated, the
boolean-valued function is evaluated, and depending on the resulting
value, either $e_1$ or $e_2$ is activated. The reactive expression $e$
terminates if the selected reactive expression terminates. Note that
the boolean-valued function is evaluated at \emph{every} instant.

\begin{figure*}[t]
\hrule
\medskip
\begin{alltt}\scriptsize
val merge          : rexp * rexp -> rexp                   (* parallel activation                      *)

val rif            : (unit -> bool) * rexp * rexp -> rexp  (* conditional activation                   *)

val halt           : unit -> rexp                          (* simply stop at every instant             *)
val nothing        : unit -> rexp                          (* terminate immediately                    *)
val loop           : rexp -> rexp                          (* reactivate rexp upon termination         *)
val terminate      : (unit -> bool) * rexp -> rexp         (* activate rexp if condition is false      *)
val init           : (unit -> unit) * rexp -> rexp         (* call fn before every activation of rexp  *)
val await          : (unit -> bool) * rexp -> rexp         (* block activation of rexp until true      *)
val when           : (unit -> bool) * rexp -> rexp         (* activate rexp when true else stop        *)
val repeat         : int * rexp -> rexp                    (* reactivate rexp a 'n' times              *)
\end{alltt}
\hrule
\caption{Reactive combinators}
\label{f:2}
\end{figure*}

The other combinators are are defined in terms of basic reactive
expressions, and the combinators 
\texttt{merge} and \texttt{rif}. Consider for example the combinator
\texttt{loop}. It takes a reactive expression $e$ and creates a new
reactive expression with the following behavior: upon activation,
it saves a copy of $e$ at its current state, and activates a copy of
that saved expression; if that copy terminates, a new copy of the saved
expression is created and activated. We initially save a copy of $e$
and work exclusively on that copy in order not to be affected by
external activations of $e$ by other reactive expressions. The
behavior just described can be implemented as follows:
\begin{alltt}\scriptsize
fun loop (e) = let val saved_e = dup (e)
                   fun l () = (activate (dup(saved_e)); 
                               l ())
               in
                   rexp l
               end
\end{alltt}

\subsection{Preemption}

Preemption refers to the possibility for one reactive expression to
force the termination of another reactive expression executing in
``parallel'' \cite{Berry93}, i.e. in another branch of a
\texttt{merge}. This is 
achieved in our framework by the \pl{SML} exception mechanism. The
following example shows how one branch of a merge can force
termination of the whole merge expression:
\begin{alltt}\scriptsize
exception Abort
let val m_exp = 
     merge (rexp (fn ()=> (print "FIRST\bsl{}n"; 
                           stop (); 
                           raise Abort)),
            loop (rexp (fn ()=> (print "SECOND\bsl{}n"; 
                                 stop ()))))
in
    rexp (fn ()=> (activate (m_exp) handle Abort => ()))
end
\end{alltt}
Since the second branch of the \texttt{merge} is a \texttt{loop}, it
never terminates, and thus the \texttt{merge} 
would never terminate if a preemption was not performed by the first
branch. 

\subsection{Micro-instants}
\label{s:2.4}

It is sometimes necessary to consider a subdivision of the
notion of instant, to provide for a finer level of control. The following
primitives are used to manage those so-called \emph{micro-instants}:
\begin{alltt}\scriptsize
val suspend : unit -> unit
val close   : rexp -> rexp
\end{alltt}
Micro-instants are created by calling the
function \texttt{suspend} instead of \texttt{stop}. A reactive
expression that calls the function \texttt{suspend} is said to be
\emph{suspended}. A suspended reactive expression behaves just like a
stopped one, unless it is wrapped with a \texttt{close}
combinator. Upon activation, a reactive expression created via a
\texttt{close} combinator will repeatedly activate the wrapped
expression until it stops or terminates. Whereas the
\texttt{suspend} function splits instants into micro-instants, the
\texttt{close} combinator performs the dual operation of combining the
micro-instants together into a single instant. 

For example, consider the following reactive expression:
\begin{alltt}\scriptsize
close (merge (rexp (fn ()=> (print "SUSPENDING "; 
                             suspend(); 
                             print "1"; stop (); 
                             print "2")),
              rexp (fn ()=> (print "A"; stop (); 
                             print "B"))))
\end{alltt}
This reactive expression will print \texttt{SUSPENDING~A1} at the
first instant, and \texttt{2B} at the second, then terminate. The
\texttt{merge} combinator activates the first reactive expression,
which suspends after printing \texttt{SUSPENDING}, and then
activates the second reactive expression, printing \texttt{A}. At this
point, the \texttt{merge} is suspended, because its first branch
is. The \texttt{close} combinator forces the completion of the
suspended reactive expressions, and thus the first branch resumes its
activation, printing \texttt{1} and then stopping. The second branch
is already stopped, so the \texttt{merge} and the \texttt{close}
stop. At the next instant, \texttt{2B} is printed in the standard
manner, and termination follows.

There is an implicit \texttt{close} wrapping the
reactive expression to which \texttt{react} is applied. It
is therefore impossible to witness micro-instants at the level of the
application that uses the reactive expression.

A common use for \texttt{suspend} and micro-instants is to suspend the
activation of a reactive expression until information has been collected
from the activation of other reactive expressions. For example, one can
implement various broadcast communication mechanisms between reactive
expressions using micro-instants \cite{Boussinot91,Boussinot96b}.

\section{An example: A simple keypad controller}
\label{s:3}

\begin{figure*}[t]
\hrule
\medskip
\begin{alltt}\scriptsize
fun mkController (n) = 
    let exception Clear
        val num = ref (0)
        fun clear () = (num := 0; raise Clear)
        val enter_rexp = rif (kEnterPressed,
                              rexp (fn ()=> (print (Int.toString (!num));
                                             clear ())),
                              halt ())
        val clear_rexp = rif (kClearPressed, rexp (clear), halt ())
        val digit_rexp = rif (kDigitPressed,
                              rexp (fn ()=> (num:=!num*10+kDigitValue ())),
                              halt ())
        val getnum_rexp = rexp (fn ()=> (activate (repeat (n,digit_rexp));
                                         activate (halt ())))
    in
        loop (rexp (fn ()=> activate (enter_rexp || clear_rexp || getnum_rexp)
                              handle Clear => ()))
    end
\end{alltt}
\hrule
\caption{Definition of the keypad controller}
\label{f:example}
\end{figure*}

Let us now consider an example of reactive code in some detail. Interesting
applications of the reactive approach are found in the programming of
widgets for user interface toolkits, such as \pl{eXene}
\cite{Gansner93}. Intuitively, a widget is an element of a user
interface that encapsulates some behavior. Basic widgets include 
buttons and text editing fields, and menus. More complex widgets can
be build up from other widgets, like dialog boxes and so on. The
obvious advantage of encapsulating behavior inside widgets is that
they can be reused in different applications.

Programming a new widget involves selecting the widgets it will be
built from and then programming the controlling behavior of the widget
(also know as the controller), taking into account the input from its
sub-widgets and the expected interface of the widget. The key
observation is that the controller is fundamentally a reactive
system. 

Suppose we want to construct a widget representing a numeric keypad
made up of 
\begin{itemize}
\item ten push buttons for the digits;
\item a push button for CLEAR, that resets the number currently
entered to 0;
\item a push button for ENTER, that prints the number currently
entered and then resets it to 0.
\end{itemize}
Suppose moreover that we parameterize the widget with respect to an
integer $n$, representing the size of the number buffer (the number of
digits it will accept).

Here is how we could represent the behavior of the keypad using the
reactive library. We program the keypad controller as a reactive
expression. We assume a mechanism that waits for windowing system
events (such as button presses) and activates the controller if an
event concerns it. To access the state of the world, we assume we have
boolean-valued functions
\begin{alltt}\scriptsize
val kDigitPressed     : unit -> bool
val kClearPressed     : unit -> bool
val kEnterPressed     : unit -> bool
\end{alltt}
that inform us if the corresponding button has been pressed in the
current instant and a function
\begin{alltt}\scriptsize
val kDigitValue       : unit -> int
\end{alltt}
that returns the last digit pressed\footnote{No usable user interface
toolkit would
require us to access the state of the interface via such
functions. We simply abstract away from the problem of communicating
windowing system events via this artificial interface.}.

We define a reactive expression for every button of the keypad. The
full controller will simply be a merge of all these 
reactive expressions. This approach simplifies the task of adding new
buttons to the keypad. The code for the controller is given in Figure
\ref{f:example}. 

The reactive expression \texttt{enter\_rexp} corresponding to the
ENTER button is simple: At every instant, we check if ENTER has been
pressed, if so, we print the number currently entered, and then clear
it and terminate. Otherwise, we stop and wait for the next
instant. The reactive expression \texttt{clear\_rexp}
handling CLEAR is the same, except that we do not print the number
currently entered. The reactive expression \texttt{digit\_rexp}
handling a digit waits for a digit to be pressed, and computes the new
number before terminating.

Recall that the keypad is parameterized by an integer $n$ representing
the size of the number buffer. We define a reactive expression
\texttt{getnum\_rexp} that will accumulate exactly $n$ digits by
activating \texttt{digit\_rexp} exactly $n$ times and then doing
nothing for the subsequent instants after $n$ digits have been
pressed. 

All of these expressions are gathered together to form the full
controller, which is obtained by applying the function
\texttt{mkController} to an integer representing the
desired size of the number buffer. The core of the controller is the
reactive expression that activates \texttt{enter\_rexp},
\texttt{clear\_rexp} and \texttt{getnum\_rexp}
concurrently\footnote{The operator \texttt{||} is simply an infix
version of \texttt{merge}, that allows for a clearer presentation of
nested merged reactive expressions.}. Since \texttt{getnum\_rexp}
never terminates, the merged expression never terminates by
itself. However, the function \texttt{clear} is used to raise a
\texttt{Clear} exception. When ENTER or CLEAR is pressed, the
exception is raised and intercepted by the exception handler of the
main reactive expression; the expression terminates and loops,
awaiting for another button press.

Suppose we wish to add a NEG button to the keypad, that negates the
number currently entered. We need only provide a function
\texttt{kNegPressed}, and a new reactive expression to handle this
case:  
\begin{alltt}\scriptsize
val neg_rexp = rif (kNegPressed,
                    rexp (fn ()=> (num := ~(!neg))),
                    halt ())
\end{alltt}
We can then add \texttt{neg\_rexp} to the general merge of the keypad
controller. It is also possible to parametrize the controller over
the buttons and corresponding reactive expressions used to
implement them.

\section{Operational semantics}
\label{s:A}

\begin{figure*}[t]
\hrule
\medskip
\begin{center}
\begin{math}
\begin{array}{rcl}

r & \in & \mbox{ReactiveId} \\
x,y,z,f & \in & \mbox{Identifier} \\
() & \in & \mbox{Unit} = \{ () \} \\
c & \in & \mbox{Constructor} = \{ \mtt{rexp}, \mtt{merge}, \mtt{rif},
           \mtt{close} \} \\
b & \in & \mbox{Basic Value} = \{ \mtt{stop}, \mtt{suspend}, \mtt{activate},\mtt{dup} \} \\
R & \in & \mbox{ReactiveSet} = \mbox{ReactiveId}
          \stackrel{\mbox{fin}}{\longmapsto} \mbox{ConstructedValue}  \\
s & \in & \mbox{State} = \{ \actstop, \actsusp, \actend \} \\
S & \in & \mbox{StateSet} = \mbox{ReactiveId}
          \stackrel{\mbox{fin}}{\longmapsto} \mbox{State} \\
v & \in & \mbox{Value} = \mbox{Unit} \cup \mbox{ReactiveId} \cup
          \mbox{Constructor} \cup \mbox{ConstructedValue} \cup \\
 & & \mbox{BasicValue} \cup \mbox{Closure} \cup \mbox{ValPair} \\
\inner{c,v} & \in & \mbox{ConstructedValue} = \mbox{Constructor}
                      \times \mbox{Value} \\
(v_1,v_2) & \in & \mbox{ValPair} = \mbox{Value} \times \mbox{Value} \\
\fn{x}{e} & \in & \mbox{Closure} = \mbox{Identifier} \times
                  \mbox{Expression} \\ 
e & \in & \mbox{Expression} = \mbox{Value} \cup \mbox{Application}
          \cup \mbox{Identifier} \cup \mbox{ExpPair} \cup \mbox{RecExp} \\
e_1~e_2  & \in & \mbox{Application} = \mbox{Expression} \times \mbox{Expression} \\
(e_1,e_2 ) & \in & \mbox{ExpPair} = \mbox{Expression} \times \mbox{Expression} \\
\rec{f}{x}{e} & \in & \mbox{RecExp} = \mbox{Identifier} \times
                      \mbox{Expression} \times \mbox{Expression} \\
R~S & \in & \mbox{ReactiveEnvironment} = \mbox{ReactiveSet} \times
            \mbox{StateSet}  
\end{array}
\end{math}
\end{center}
\hrule
\caption{Semantic objects}
\label{f:objs}
\end{figure*}

We describe the semantics of reactive expressions in terms of a simple
functional language, in the spirit of \cite{Berry92a,Reppy92}. The
syntax of the language is given by the following grammar: 
\begin{eqnarray*}
M & = & x ~|~ () ~|~ M_1~ M_2 ~|~ (M_1,M_2) ~|~ (M_1;M_2) \\ 
 & & ~|~ \fn{x}{M} ~|~ \rec{f}{x}{M} 
\end{eqnarray*}
where $x$ and $f$ are alphabetic identifiers. The semantic
objects are given in Figure \ref{f:objs}. As in
\cite{Berry92a}, the set of expressions is a superset of both the set
of values and the set of lexical phrases. The set of identifiers
includes all possible alphabetic identifiers, including the
constructors and basic values. The constructors and basic values
define the reactive behavior of the language. They have no special
syntax beyond their existence as identifiers. The notation
$\stackrel{\mbox{fin}}{\longmapsto}$ is used to denote a finite mapping.

A reactive expression is tagged with a unique reactive ID:
whenever a new reactive expression is created via \texttt{rexp} or by
applying a combinator to existing reactive 
expressions, a new reactive ID is allocated and associated to the
reactive expression. A reactive environment $R~S$ is used to store
information relating to reactive expressions. The map $R$ holds the
bindings between reactive IDs and actual reactive expressions, stored
as constructed values. The map $S$ stores the current state of
reactive expressions, indexed by reactive ID. The state of a reactive
expression is either stopped, suspended or terminated.  Given $M$ a
finite map, we use the notation $M[m:v]$ to denote the new map defined
by:
\[ M[m:v] (m') = \left\{ \begin{array}{ll}
                           M(m') & \mbox{if $m'\not=m$} \\
                           v     & \mbox{if $m'=m$}
                         \end{array} \right. \]

The semantics is described used Plotkin's Structural Operational
Semantics \cite{Plotkin81}, and extends the semantics of \pl{Reactive
C} given in \cite{Boussinot92a}.

The semantics we describe has two levels. The
first level is a semantics for the core language, given in Figure
\ref{f:core}, expressed as a 
conventional reduction rule semantics. The semantics of the core
language is 
complicated by the fact that expressions can occur in two
contexts: the normal sequential context, and in the context of a
reactive expression (captured by a \texttt{rexp} constructor). The
reduction relation
\[ \at{e}{R}{S}{e'}{R'}{S'} \]
denotes the reduction of expression $e$ into expression $e'$, possibly
transforming the reactive environment $R~S$ into $R'~S'$. The
reduction is labeled by an action $\alpha$: \texttt{END} if the
reduction terminates instantly, \texttt{STOP} if the reduction is
stopped and \texttt{SUSP} if the reduction is suspended. Stopped and
suspended reductions can only occur within the context of
\texttt{rexp} constructed expression. 

\begin{figure*}[t]
\hrule
\medskip
{\footnotesize
\begin{center}
\begin{math}
\begin{array}{cc}

\Rule{\at{e_1}{R}{S}
          {e_1'}{R'}{S'}}
     {\at{e_1~e_2}{R}{S}
          {e_1'~e_2}{R'}{S'}} &

\Rule{\at{e}{R}{S}
          {e'}{R'}{S'}}
     {\at{v~e}{R}{S}
          {v~e'}{R'}{S'}} \\[5ex]

\Rule{\at{e_1}{R}{S}
          {e_1'}{R'}{S'}}
     {\at{(e_1,e_2)}{R}{S}
          {(e_1',e_2)}{R'}{S'}} &

\Rule{\at{e}{R}{S}
          {e'}{R'}{S'}}
     {\at{(v,e)}{R}{S}
          {(v,e')}{R'}{S'}} \\[5ex]

\Rule{\at{e_1}{R}{S}
          {e_1'}{R'}{S'}}
     {\at{(e_1;e_2)}{R}{S}
          {(e_1';e_2)}{R'}{S'}} &

\Rule{\at{e_1}{R}{S}
          {v}{R'}{S'}}
     {\at{(e_1;e_2)}{R}{S}
          {e_2}{R'}{S'}} \\[5ex]

\multicolumn{2}{c}{
\Axiom{\et{\fn{x}{e}~v}{R}{S}
           {e\{v/x\}}{R}{S}} 
} \\[5ex]

\multicolumn{2}{c}{
\Axiom{\et{\rec{f}{x}{e}}{R}{S}
           {\fn{x}{e\{\rec{f}{x}{e}/f\}}}{R}{S}}
} \\[5ex]

\multicolumn{2}{c}{
\Rule{r\not\in \mbox{dom}(R)}
     {\et{c~v}{R}{S}
         {r}{R[r:\inner{c,v}]}{S[r:\actstop]}}
} \\[5ex]

\Rule{\ett{r}{R}{S}
          {R'}{S'}}
     {\et{\mtt{activate} (r)}{R}{S}
            {()}{R'}{S'}} &

\Rule{\att{r}{R}{S}
          {R'}{S'} \transp
      \alpha\not=\actend}
     {\at{\mtt{activate} (r)}{R}{S}
         {\mtt{activate} (r)}{R'}{S'}} \\[5ex]

\Axiom{\st{\mtt{stop} ()}{R}{S}
          {()}{R}{S}}  & 

\Axiom{\ut{\mtt{suspend} ()}{R}{S}
          {()}{R}{S}} \\[5ex]

\multicolumn{2}{c}{
\Rule{R(r) = \inner{c,v} \transp
      S(r) = s \transp
      r' \not\in R}
     {\et{\mtt{dup} (r)}{R}{S}
         {r'}{R[r':\inner{c,v}]}{S[r':s]}}
} \\[5ex]

\Rule{S(r)=\actend}
     {\ett{r}{R}{S}
          {R}{S}} & 

\Rule{S(r)\not=\actend \transp
      \atr{r}{R}{S}
          {R'}{S'}}
     {\att{r}{R}{S}
          {R'}{S'}} 
\end{array}
\end{math}
\end{center}
}
\hrule
\caption{Semantics of core language}
\label{f:core}
\end{figure*}

As we already noted, the only reductions allowed for the core language in
the normal sequential context are terminated reductions (labeled
\texttt{END}). The function \texttt{react} is the only function that
may be called from the sequential context. We do not give the
semantics of \texttt{react} to unclutter the presentation of the
rules, but \texttt{react} behaves as \texttt{activate} in the
sequential context --- returning \texttt{true} or \texttt{false}
depending on the resulting status of the reactive expression to which
it is applied. 

In the context of a reactive expression, the core language is allowed
the full range of labeled reductions. The basic functions
\texttt{stop}, \texttt{suspend}, \texttt{activate} and \texttt{reset}
are allowed in such a context.

The interesting rules in that part of the semantics are the rules for
\texttt{activate}, which acts upon reactive expressions (really
reactive IDs representing reactive expressions). The intuitive
interpretation of \texttt{activate} is to activate the given reactive
expression, which propagates the activation according to the 
structure of the reactive expression.

The rules for \texttt{activate} involve the reduction relation
\[ \att{r}{R}{S}{R'}{S'} \]
that act on a reactive expression whose ID is $r$. Again, the
reduction is labeled by an action $\alpha$ indicating if the
reduction is terminated, stopped or suspended. The intuition behind
this reduction relation is that if the reactive expression denoted by
$r$ is stopped, then the activation is not propagated, and
\texttt{activate} returns immediately. If the reaction is not
terminated, then the activation is 
propagated to the reactive expression via the reduction
\aredr. The reduction relation \aredt{} is not formally necessary, but
does greatly simplify the semantic rules.

The actual activation of reactive expressions is expressed by the
reduction relation 
\[ \atr{r}{R}{S}{R'}{S'} \]
denoting the activation of the reactive expression whose ID is $r$. The
rule to apply depends on the structure of the reactive expression,
whose constructed value is extracted from the reactive
environment. Again, this reduction rule is labeled by an action
$\alpha$ indicating if the reaction terminates, stops or
suspends. Figure \ref{f:rexp} gives the semantics of reactive
expressions. If the reactive expression is a \texttt{rexp}
constructed value, the reduction of the expression involves the
reduction of an expression in the core language via \ared{} transitions.

The semantics of the \texttt{merge} combinator use a function $\star$
on actions, defined as follows:
\begin{center}
\begin{math}
\begin{array}{|c|ccc|}
\hline
\star & \actsusp & \actstop & \actend \\ \hline
\actsusp & \actsusp & \actsusp & \actsusp \\
\actstop & \actsusp & \actstop & \actstop \\
\actend & \actsusp & \actstop & \actend \\
\hline
\end{array}
\end{math}
\end{center}

\begin{figure*}[!t]
\hrule
\medskip
{\footnotesize
\begin{center}
\begin{math}
\begin{array}{c}

\Rule{R(r)=\cons{rexp}{v_1} \transp
      \at{v_1~()}{R}{S}
         {e}{R'}{S'} \transp
      \alpha\not=\actend}
     {\atr{r}{R}{S}
          {R'[r:\cons{rexp}{\fn{()}{e}}]}{S'[r:\alpha]}} \\[5ex]

\Rule{R(r)=\cons{rexp}{v_1} \transp
      \et{v_1~()}{R}{S}
         {v}{R'}{S'}}
     {\etr{r}{R}{S}
          {R'}{S'[r:\actend]}} \\[5ex]

\Rule{R(r)=\cons{merge}{(r_1,r_2)} \transp
      S(r)\not=\actsusp \transp
      \tr{r_1}{R}{S}{$\alpha_1$}
         {R'}{S'} \transp
      \tr{r_2}{R'}{S'}{$\alpha_1$}
         {R''}{S''}}
     {\tr{r}{R}{S}{$\alpha_1\star\alpha_2$}
         {R''}{S''[r:\alpha_1\star\alpha_2]}} \\[5ex]

\Rule{R(r)=\cons{merge}{(r_1,r_2)} \transp
      S(r_1)=S(r_2)=\actsusp \transp
      \tr{r_1}{R}{S}{$\alpha_1$}
         {R'}{S'} \transp
      \tr{r_2}{R'}{S'}{$\alpha_1$}
         {R''}{S''}}
     {\tr{r}{R}{S}{$\alpha_1\star\alpha_2$}
         {R''}{S''[r:\alpha_1\star\alpha_2]}} \\[5ex]

\Rule{R(r)=\cons{merge}{(r_1,r_2)} \transp
      S(r_1)=\actsusp \transp
      S(r_2)\not=\actsusp \transp
      \tr{r_1}{R}{S}{$\alpha$}
         {R'}{S'} \transp}
     {\tr{r}{R}{S}{$\alpha\star S(r_2)$}
         {R'}{S'[r:\alpha\star S(r_2)]}} \\[5ex]

\Rule{R(r)=\cons{merge}{(r_1,r_2)} \transp
      S(r_1)\not=\actsusp \transp
      S(r_2)=\actsusp \transp
      \tr{r_2}{R}{S}{$\alpha$}
         {R'}{S'} \transp}
     {\tr{r}{R}{S}{$S(r_1)\star\alpha$}
         {R'}{S'[r:S(r_1)\star\alpha]}} \\[5ex]

\Rule{R(r)=\cons{rif}{(v,r_1,r_2)} \transp
      S(r)\not=\actsusp \transp
      \et{v~()}{R}{S}
         {\true}{R'}{S'} \transp
      \atr{r_1}{R'}{S'}
          {R''}{S''}}
     {\atr{r}{R}{S}
          {R''}{S''[r:\alpha]}} \\[5ex]

\Rule{R(r)=\cons{rif}{(v,r_1,r_2)} \transp
      S(r)\not=\actsusp \transp
      \et{v~()}{R}{S}
         {\false}{R'}{S'} \transp
      \atr{r_2}{R'}{S'}
          {R''}{S''}}
     {\atr{r}{R}{S}
          {R''}{S''[r:\alpha]}} \\[5ex]

\Rule{R(r)=\cons{rif}{(v,r_1,-)} \transp
      S(r_1)=\actsusp \transp
      \atr{r_1}{R}{S}
          {R'}{S'}}
     {\atr{r}{R}{S}
          {R'}{S'[r:\alpha]}} \\[5ex]

\Rule{R(r)=\cons{rif}{(v,-,r_2)} \transp
      S(r_2)=\actsusp \transp
      \atr{r_2}{R}{S}
          {R'}{S'}}
     {\atr{r}{R}{S}
          {R'}{S'}} \\[5ex]

\Rule{R(r)=\cons{close}{r'} \transp
      \atr{r'}{R}{S}
          {R'}{S'} \transp
      \alpha\not=\actsusp}
     {\atr{r}{R}{S}
          {R'}{S'[r:\alpha]}} \\[5ex]

\Rule{R(r)=\cons{close}{r'} \transp
      \utr{r'}{R}{S}
          {R'}{S'} \transp
      \atr{r}{R'}{S'}
          {R''}{S''}}
     {\atr{r}{R}{S}
          {R''}{S''}}

\end{array}
\end{math}
\end{center}
}
\hrule
\caption{Semantics of reactive expressions}
\label{f:rexp}
\end{figure*}

\subsection{Implementation}
\label{s:B}

The implementation of the library is a direct translation of the
operational semantics. The core of of the implementation is
a function \texttt{step} that plays the role of the \redr{} transition
in the semantics. It is used to activate a reactive expression and
returns the state of the expression after the activation. Every basic
combinator may be expressed by the \texttt{step} function and the
basic reactive expression constructor, by simply defining it via its
semantic reduction rules. We therefore only need to concentrate on the
implementation of \texttt{step}, \texttt{stop} and
\texttt{suspend}. 

The library was implemented with the \pl{Standard ML of New Jersey}
compiler \cite{Appel87}. The compiler provides
\texttt{callcc} \cite{Harper93}, an extension to \pl{SML} that allows
the expression of powerful control abstraction in a typed setting. It
lets one grab the current continuation of the evaluation of an
expression as a first-class object and resume it at will. 

A reactive expression is implemented as a tuple containing the 
continuation of the reactive expression and the current state of the
expression. Calling \texttt{step} on the tuple simply throws the
stored continuation to resume the evaluation of the expression, after
saving the current continuation. This latter continuation will be
thrown if \texttt{stop} is called from the reactive expression
code. The function \texttt{stop} saves the current continuation of the
reactive expression in the tuple, and throws the continuation saved by
the \texttt{step} function, resuming the evaluation of the code
calling \texttt{step}. The function \texttt{suspend} is similarly
implemented. The technique used is analogous to the one used by Wand
\cite{Wand80} and Reppy \cite{Reppy92} to implement concurrent threads
via continuations. 

\section{Comparison with other reactive frameworks}
\label{s:4}

The library described in this paper evolved from a desire to port the
reactive framework of \pl{Reactive C} \cite{Boussinot91} to the
higher-order language \pl{SML}. It is instructive to compare our
system against both the original  
\pl{Reactive C} and its derivative, the \pl{Java} toolkit
\pl{SugarCubes} \cite{UNSTABLE:Boussinot97}. We also compare the
library to various other frameworks for programming reactive systems.

\subsection{\pl{Reactive C} and \pl{SugarCubes}}

The principal difference between the formalism in this paper and the
formalisms of both \pl{Reactive C} and \pl{SugarCubes} relates to the
programming paradigm embodied by the underlying languages. The
\pl{Reactive C} formalism extends the imperative language \pl{C}
\cite{Kernighan88} where programs are viewed as sequence of
commands. The formalism defines a ``machine'' executing a sequence of
reactive ``instructions''. The \pl{SugarCubes} toolkit extends the
object-oriented language \pl{Java} \cite{Gosling96}, and also uses the same
imperative approach.

To illustrate the differences between ``reactive instructions'' and
reactive expressions as we defined them in this paper, observe that
our framework can be expressed as a datatype. Following \pl{SML}'s
notation, one may define the following:
\begin{alltt}\scriptsize
datatype rexp = REXP of unit -> unit
              | MERGE of rexp * rexp
              | RIF of (unit -> unit) * rexp * rexp
              | ...
\end{alltt}
A reactive expression becomes a tree-shaped data structure, and
\texttt{react} simply walks the given tree. In fact, the semantics
of reactive expressions given in Section \ref{s:A} uses
exactly this view of reactive expressions as a constructed
datatype. In this framework, the leaves of the structure are basic  
reactive expressions that contain arbitrary \texttt{stop},
\texttt{suspend} and \texttt{activate} calls. Sample reactive code
would look like:
\begin{alltt}\scriptsize
MERGE (REXP (fn ()=> (print "1"; stop(); print "2")),
       REXP (fn ()=> (print "A"; stop(); print "B")))
\end{alltt}

If we were to implement the reactive ``instructions'' approach in
\pl{SML} via a datatype description as above, we would obtain
something like the following:
\begin{alltt}\scriptsize
datatype rinst = EXP of unit -> unit
               | STOP
               | SUSPEND
               | ACTIVATE of rinst
               | SEQUENCE of rinst list
               | MERGE of rinst
               | RIF of (unit -> unit) * rinst * rinst
               | ...
\end{alltt}
We do not allow arbitrary calls to \texttt{stop} and \texttt{suspend}
in basic expressions. Rather, the end of instants are explicitly
specified in the datatype. Basic expressions always terminate
immediately. This means that much of the structure that in
our framework would fit in a basic reactive expression
\texttt{REXP} now needs to be explicitly added to the datatype (for
example, a way to describe sequences of reactive instructions). The
sample code given above would now look like:
\begin{alltt}\scriptsize
MERGE (SEQUENCE [EXP (fn ()=> print "1"), STOP, 
                 EXP (fn ()=> print "2")],
       SEQUENCE [EXP (fn ()=> print "A"), STOP, 
                 EXP (fn ()=> print "B")])
\end{alltt}
A reactive library implemented via reactive instructions (using the
model of SugarCubes) is part of the Standard ML of New Jersey
Library\footnote{J. Reppy, Personal communication, 1997.}.

The first approach, which we followed in this paper, allows
for a clearer syntax, by directly using \pl{SML} control-flow primitives
(sequencing, local declarations) which need to be redefined in
the datatype for the second approach. Moreover, the first approach
allows one to easily reuse existing higher-order functions in a
reactive way. One can easily write:
\begin{alltt}\scriptsize
REXP (fn ()=> app (stop o print) ["1","2","3","4"] )
\end{alltt}
which prints one number of the list at every instant until
termination. Expressing this reactive expression in the second
approach seems difficult. On the other hand, the reactive code in the
second approach is easier to analyze (for the purpose of compilation, for
example), since the end of instants is fully characterized by
the actual data structure representing the reactive code --- there is
no need to analyze the control-flow of an arbitrary \pl{SML} expression
calling \texttt{stop} and \texttt{suspend}.

\subsection{Synchronous languages}

Synchronous languages are among the most popular languages for
programming reactive systems. These include \pl{Esterel}
\cite{Berry92}, \pl{Lustre} \cite{Caspi87} and \pl{Signal}
\cite{LeGuernic86}. These languages are all based on the same notions of instants
and activations that we describe in this paper, but with important
additions. In the case of \pl{Esterel}, we have the following:
\begin{enumerate}
\item The instants are assumed to take zero time and are atomic. This
is the \emph{synchrony hypothesis}. 
\item Communication between parallel reactions is done via broadcast
signals, and is instantaneous.
\item Preemption can be triggered by the presence of a specified
signal in the instant under consideration. 
\end{enumerate}

These characteristics allow the code for \pl{Esterel} (and synchronous
languages in general) to be efficiently compiled into a finite-state
automaton, which can be translated into a program in a sequential
language. There exists a translator taking the output of the 
\pl{Esterel} compiler into \pl{SML} code implementing the
corresponding finite-state automaton\footnote{J. Riecke, Personal
communication, 1997.}.

Our framework does not support the synchrony hypothesis of synchronous
languages, and provides no communication mechanism between various
parallel reactions beyond shared memory. As such, it does not support
compilation into finite-state machines, and can be considered
lower-level than synchronous languages. Boussinot and de Simone
showed in \cite{Boussinot96a} that it is possible to translate a
synchronous language into a framework similar to the one described in
this paper. This justifies our intended goal of using the library as a
target language for experimental extensions to synchronous languages,
such as higher-order synchronous languages. These extensions might not
preserve finite-state semantics, but may still be useful as a
convenient notation for various types of processes.

Higher-order extensions to synchronous languages include the work of
Caspi and Pouzet \cite{Caspi96} on extending the dataflow
synchronous language Lustre with higher-order functions.

\subsection{\pl{Fran}}

The \pl{Fran} system \cite{Elliott97} is a 
reactive framework for programming  multimedia animations in
\pl{Haskell} \cite{Peterson97}. It defines the notions of behaviors
and events to program reactive animations. A behavior is fundamentally
a function of time, and it is possible to specify behaviors with
respect to events. The principal difference between the \pl{Fran} approach
and the one in this paper is that \pl{Fran} is based on a continuous
time model as opposed to our discrete time model divided into instants.

\subsection{Coroutine facilities}

Languages providing facilities for defining coroutines can be used to
define a reactive framework such as the one we present in this
paper. For example, the programming language \pl{Icon}
\cite{Griswold96} provides \emph{co-expressions}, which are
expressions that can be suspended and resumed at a later
time. When it suspends, a co-expression needs to state to which other
co-expression it is relinquishing control. Our basic notions of
activation and instants can be viewed as a hierarchical use of
co-expressions.

\section{Future work}
\label{s:5}

The interesting questions about the framework presented in this
paper all relate to the interaction between the reactive formalism and
the underlying mostly-functional approach of \pl{SML}. For example,
the current implementation of the \texttt{rif} combinator uses a
\texttt{(unit -> unit)} function to be evaluated at every instant to
determine which branch of the reactive conditional to activate.  This
means that any external value used by the test must be a
reference.  

Possible extensions to the framework include adding parameters to
\texttt{react} that will be propagated through every combinator and
used by \texttt{rif} during the evaluation of the conditional
test. Dually, we can give a return type to reactive
expressions, so that a value may be returned when a reaction stops or
terminates. One should then augment the \texttt{merge} combinator with
a function specifying how to combine the values returned by the two
branches.   

Futher investigations into the interaction between reactivity and
higher-order functions will involve the implementation of the reactive
framework as a monad \cite{Wadler95} in the purely functional language
Haskell. Work by Claessen \cite{UNSTABLE:Claessen97} on expressing
concurrency as a monad via explicitly interleaved atomic actions
closely follow our \texttt{merge} combinator. It would also be of
interest to embed the reactive framework in a typed
$\lambda$-calculus, in a way similar to the semantics of reactivity
given in terms of a process calculus in \cite{Boussinot93}.

\section{Conclusion}

We have described in this paper a reactive library for \pl{SML} that
implements the reactive paradigm exemplified by modern
languages such as \pl{Esterel}. The library provides primitives that
capture the essence of the reactive paradigm, namely the notions of
instants and activations. The library is intended to be a low-level
system upon which more sophisticated reactive behavior can be built,
providing a convenient framework for prototyping various higher-level
reactive languages.

\section*{Acknowledgments}

I wish to thank Fr\'ed\'eric Boussinot and John Reppy for careful
readings of the manuscript, as well as Nevin Heintze, Hormoz
Pirzadeh, Jon Riecke, Michael Sperber, Walid Taha and the anonymous
referees for useful comments on various aspects of this work. 

{\footnotesize
\bibliography{main,unstable}
}

\end{document}